\documentclass[reprint,amsmath,amssymb,aps,pre,article]{revtex4-2}
\usepackage{color}
\usepackage{graphicx}
\usepackage{dcolumn}
\usepackage{bm}
\usepackage[mathlines]{lineno}
\usepackage{amsfonts}
\usepackage{amsmath}
\usepackage{sidecap}
\usepackage{float}
\usepackage{parskip}
\usepackage{grffile}

\usepackage{color}
\usepackage{hyperref}
\usepackage{breakurl}
\usepackage{hhline}
\usepackage{mathtools}
\usepackage{comment}
\usepackage[normalem]{ulem}
\usepackage{url}
\usepackage{hyperref}

\begin{document}
\preprint{APS/123-QED}

\title{Automated discovery of dimensional reductions in coupled oscillators}

\author{Dishant Sisodia}
\email{dishantsisodia.physics@gmail.com}

\author{Sarika Jalan} \email{sarikajalan9@gmail.com: Corresponding author}
\affiliation{Complex Systems Lab, Department of Physics, Indian Institute of Technology Indore, Khandwa Road, Simrol, Indore-453552, India}

\begin{abstract}
There have been persistent efforts to uncover low-dimensional macroscopic dynamics in large systems of interacting microscopic units, in which underlying physical principles govern their collapse onto low-dimensional manifolds. Here, focusing on interacting phase oscillator systems, we show that an unsupervised autoencoder framework can determine their intrinsic dimensionality, in agreement with the predictions of Watanabe–Strogatz theory. Systematic sweeps over the latent dimensions identify the analytically predicted reduced dimensions, while supervised regression confirms that the learned latent variables capture the full information contained in the theoretical Watanabe–Strogatz coordinates. We demonstrate this capability for two prototype systems: Kuramoto oscillators and resistively loaded Josephson junctions. We further establish the diagnostic specificity of the framework by showing that systems lacking such reducible dynamics do not exhibit a sharp dimensional reduction.

\end{abstract} 
\maketitle
\section{Introduction}
In recent years, the intersection of machine learning and nonlinear dynamics has emerged as a powerful paradigm for analyzing complex systems \cite{10.1063/5.0016505, Tsironis2025}. Deep learning algorithms are now routinely deployed to discover governing physical equations \cite{Brunton2016, doi:10.1073/pnas.1906995116, Udrescu2020AIFeynman}, construct physics-informed neural network architectures \cite{RAISSI2019686}, and forecast chaotic time series while reconstructing attractors \cite{kong2023reservoir, PhysRevResearch.6.043194, Jpathak, yadav2025predicting}. Concurrently, significant momentum has grown around two complementary frontiers. On one hand, analytical tools and principles from physics are increasingly applied to demystify the internal mechanisms and representations of learning models \cite{PhysRevE.110.034211, 8b7z-fnjd, kantamneni2025, sisodia2026finite, PhysRevLett.132.057301}. On the other, researchers are investigating whether machine learning can autonomously extract fundamental physical laws directly from observational data; spanning machine assisted discovery of physical concepts \cite{PhysRevLett.124.010508}, classification of phase transitions \cite{PhysRevE.96.022140, 2v76-dmg6}, and automated symmetry identification \cite{PhysRevLett.126.180604, PhysRevLett.128.180201}.

Despite these advances, a fundamental question remains: can machine learning autonomously discover exact, reduced macroscopic dynamics directly from high-dimensional microscopic observations? Addressing this is essential, since such low-dimensional reductions allow complex collective behaviors to be analyzed directly, making the underlying dynamics exact and solvable. Coupled phase oscillators provide an ideal setting for this challenge, governing phenomena from pedestrian synchrony \cite{strogatz2005crowd} to firefly synchronization \cite{doi:10.1086/415929}, with the Kuramoto model serving as the foundational paradigm \cite{10.1007/BFb0013365}. Originally motivated by the dynamics of Josephson junction arrays \cite{PhysRevLett.68.2265}, Watanabe and Strogatz (WS) uncovered an exact coordinate transformation that reduces the high-dimensional microscopic states to a minimal set of macroscopic variables while fully preserving the dynamical information of the original system \cite{WATANABE1994197}. This reduction applies broadly to globally coupled identical phase oscillators \cite{10.1063/1.3247089, 10.1063/5.0283600}, establishing these systems as an optimal testbed for data-driven dimensional reduction. To determine whether deep learning architectures can extract such closed form macroscopic dynamics, autoencoders and their specialized variants (\cite{Kingma_2019}, \cite{NIPS2007_c60d060b}) offer a promising framework. By enforcing a low-dimensional bottleneck, autoencoders compress high-dimensional observational trajectories into compact latent representations. Few existing autoencoder studies in nonlinear dynamical systems have demonstrated success in reconstructing dynamical trajectories from observational data \cite{10.1063/5.0232584}, estimating effective manifold dimensions in partial differential equations and multi variable chaotic systems \cite{Zeng_2024, 10.1063/5.0055673}, and characterizing topological flows \cite{10.1063/5.0013714}. However, the ability of an autoencoder's latent space to uncover exact underlying physical symmetries remains unexplored in prior work.

In this work, we use autoencoders to identify the intrinsic dimensionality of coupled oscillator systems directly from the phase trajectories and show that it agrees with the predictions of WS theory. Systematic sweep over the latent-dimension accurately recovers the theoretical macroscopic dimensions, while supervised regression confirms that the learned representations retain the full dynamical information of the WS coordinates through a smooth, nonlinear parametrization. Testing our framework on Kuramoto oscillators, Josephson junction arrays, and non-reducible heterogeneous control systems demonstrates that autoencoders can reliably discover minimal low-dimensional dynamics directly from the phase trajectories. We further extend this approach to multi-population dynamical systems exhibiting chimera states and contrarian–conformist dynamics. 
\begin{figure*}[t!]
\centering
\includegraphics[scale=0.4, trim= 2cm 6cm 0 5cm, clip]{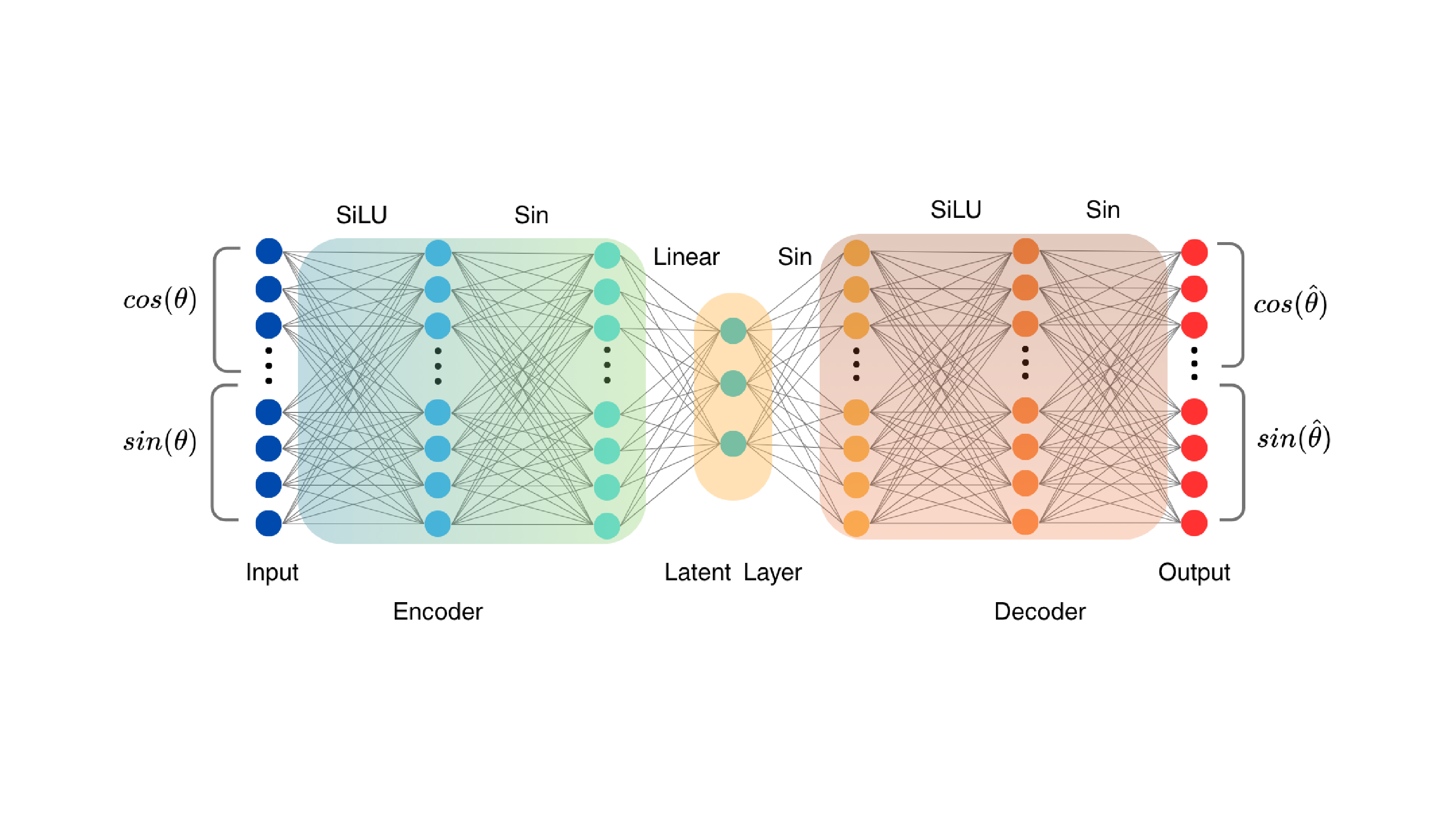}
\caption{Schematic of a fully connected autoencoder. Phase trajectories of $N$ coupled oscillators are embedded as $2N$-dimensional vectors comprising $[\cos\theta, \sin\theta]$. Both the encoder and decoder consist of two hidden layers with layer-specific activation functions (Sin/SiLU) labeled above. The latent bottleneck dimension $d_z$ is systematically varied across training runs to evaluate the reduced manifold dimensionality.}
\label{fig1}
\end{figure*}
\section{Autoencoder architecture}
An autoencoder is a neural network architecture designed to perform unsupervised dimensionality reduction by extracting essential features from complex data \cite{10.1126/science.1127647}. The model comprises two main functional mappings: an encoder, $E_{\theta}$, which maps the input signal into a reduced latent space $\mathbf{z}$, and a decoder, $D_{\phi}$, which reconstructs the input vector from this compressed representation. The intermediate bottleneck layer forces the network to retain only the most crucial attributes of the underlying data distribution. Mechanically, the network processes information across $L$ sequential layers. The output state of layer $l$ having $n_l$ neurons, denoted as $\mathbf{x}^{(l)} \in \mathbb{R}^{n_l}$, is computed by applying an element-wise activation function to the preceding layer's output:
\begin{equation}
    \mathbf{x}^{(l)} = \sigma_l \left( \mathbf{W}_l \mathbf{x}^{(l-1)} + \mathbf{b}_l \right), \nonumber
\end{equation}
where $\mathbf{W}_l \in \mathbb{R}^{n_l \times n_{l-1}}$ and $\mathbf{b}_l \in \mathbb{R}^{n_l}$ represent the trainable weight matrix and bias vector for layer $l$, respectively, and $\sigma_l(\cdot)$ provides the non-linear activation. Although, several activation functions exist in the literature, we use \textit{Sin} and Sigmoid Linear Unit (\textit{SiLU}) activations, where $SiLU(x) = \frac{x}{1 + e^{-x}}$.
Since the primary task is to learn a compressed representation that faithfully preserves the input dynamics, the model parameters are trained by minimizing the mean squared error (MSE) reconstruction loss:
\begin{equation}
    \mathcal{L}(\theta, \phi) = \frac{1}{T} \sum_{i=1}^{T} \left\| \mathbf{x}_i - \hat{\mathbf{x}}_i \right\|_2^2, \nonumber
\end{equation}
where $T$ denotes the total number of training instances, $\mathbf{x}_i$ represents the original input vector, and $\hat{\mathbf{x}}_i = D_{\phi}(E_{\theta}(\mathbf{x}_i))$ is the reconstructed output from the autoencoder. The joint parameter set $\Omega = \{\mathbf{W}_l, \mathbf{b}_l\}_{l=1}^{L}$, comprising all trainable weight matrices and bias vectors across the encoder ($E_\theta$) and decoder ($D_\phi$) networks, is optimized by minimizing the loss function via standard backpropagation methods.

As depicted in the schematic (Fig.~\ref{fig1}), we construct a fully connected feedforward autoencoder with a tunable latent dimension $d_z$, enabling systematic identification of the system's low-dimensional invariant space. To resolve the boundary discontinuities associated with angular variables $\theta_j \in (-\pi, \pi)$ due to $2\pi$-periodicity across $N$ coupled units, the autoencoder receives a $2N$-dimensional input vector composed of concatenated trigonometric features $[\cos\theta_j(t), \sin\theta_j(t)]_{j=1}^N$. The encoder transforms this input through a first hidden layer employing a \textit{SiLU} activation, followed by a second hidden layer conditioned by a \textit{Sin} activation function. A direct linear projection ($\sigma(x) = x$) is then applied to the second hidden layer to form the unconstrained bottleneck latent variables $\mathbf{z}$. Reversing this architecture, the decoder processes the latent representations through a first hidden layer with a \textit{Sin} activation, a second hidden layer with a \textit{SiLU} activation, and an output layer that reconstructs the trigonometric states $[\cos\hat{\theta}_j(t), \sin\hat{\theta}_j(t)]_{j=1}^N$ via a final \textit{Sin} activation function.
All the layers except the latent layer, have the same number of $2N$ neurons.

\section{Coupled oscillator models and Watanabe-Strogatz theory}
\begin{figure}[t!]
    \centering
    \includegraphics[width = \columnwidth]{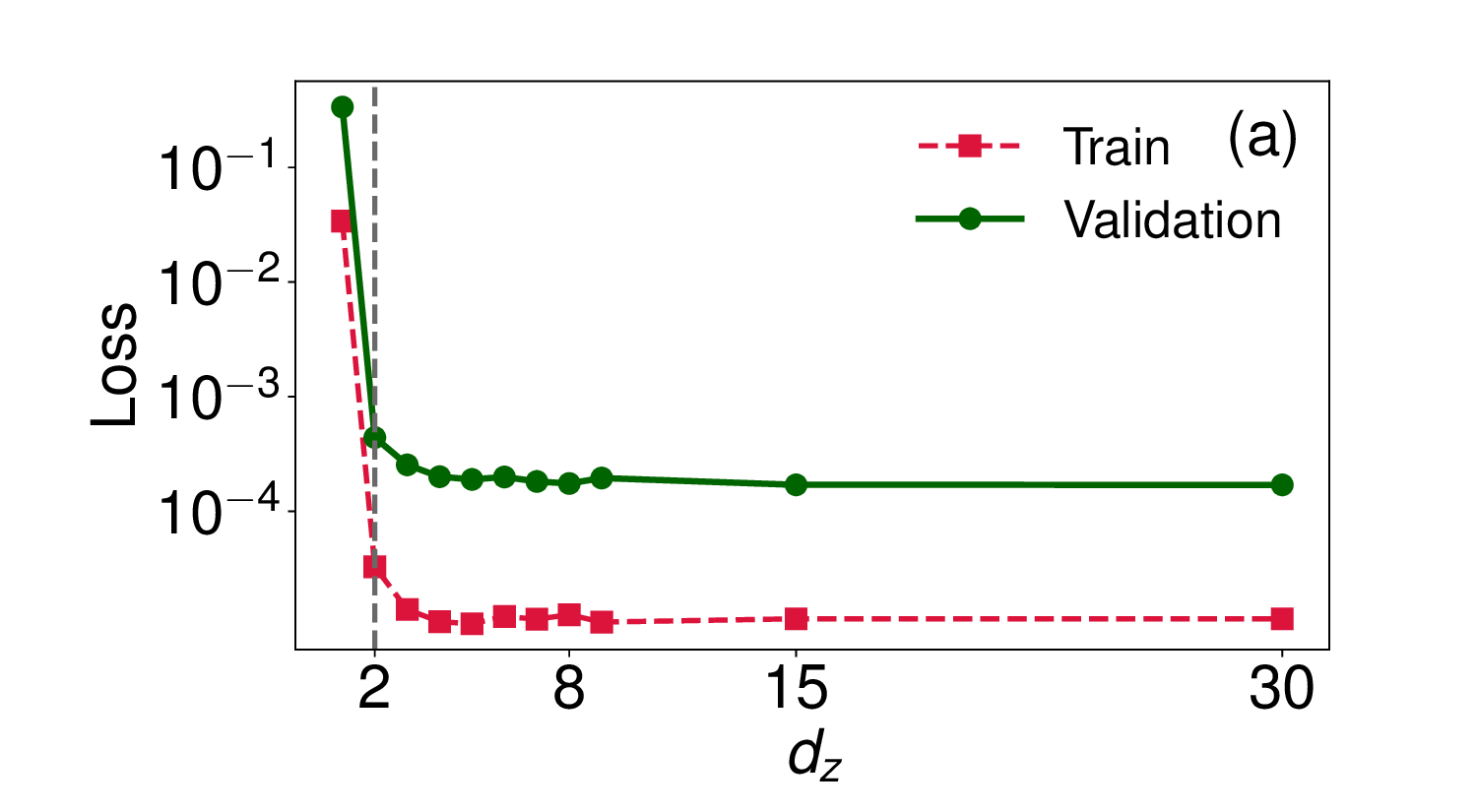} \\
    \includegraphics[width = \columnwidth]{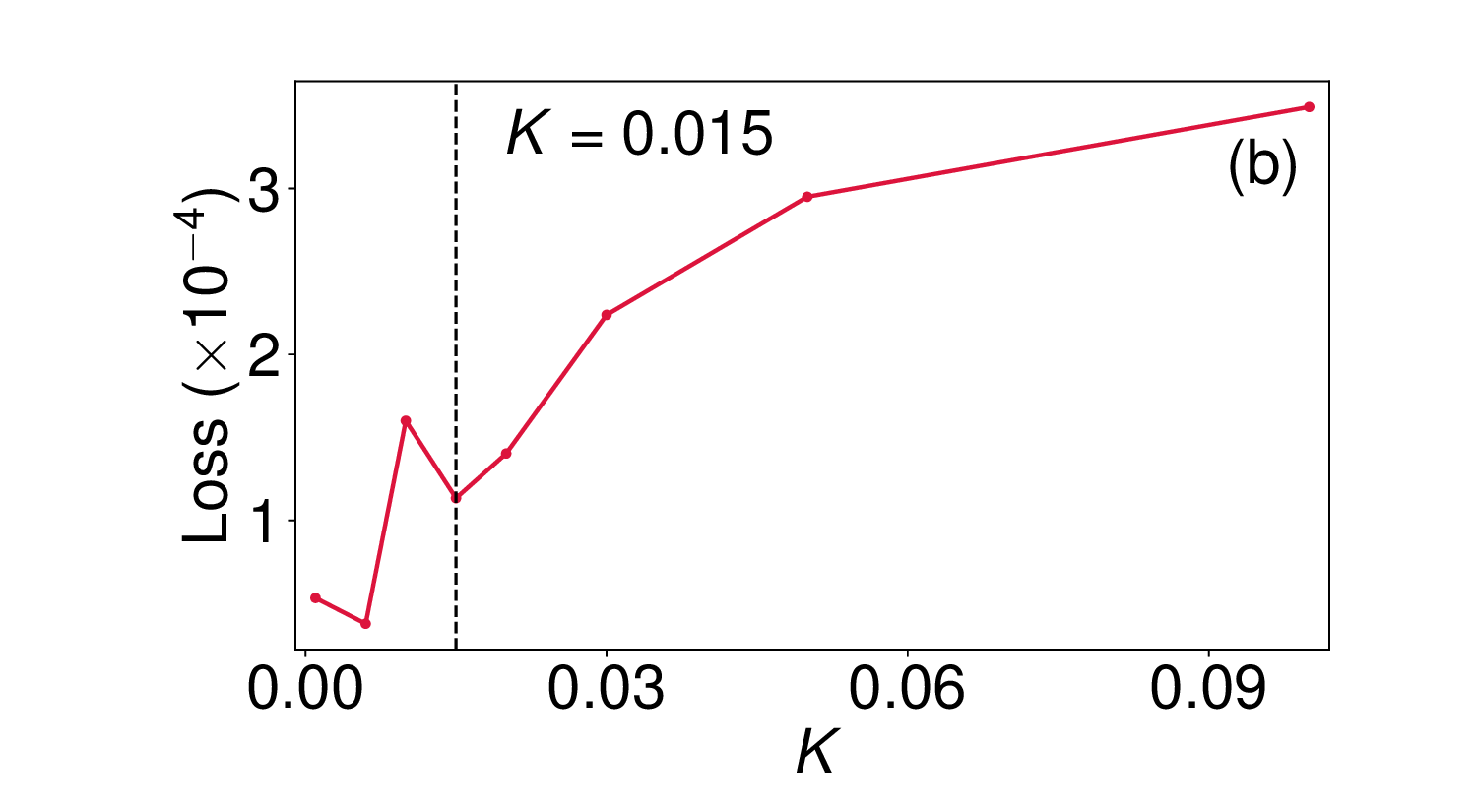}  
    \caption{Identification of intrinsic dimensionality for $N = 64$ Kuramoto oscillators ($\omega = 0.4$) trained at coupling strength $K = 0.015$. (a) Training and validation MSE as a function of latent dimension $d_z$, exhibiting a sharp elbow at $d_z = 2$ in exact agreement with the WS theoretical prediction. (b) MSE evaluated on unseen trajectories across varying coupling strengths $K$. The vertical dashed line marks the training parameter ($K = 0.015$), demonstrating generalization across different dynamical regimes. Training is performed with $B = 256$, $N_{\text{epochs}} = 350$}.
    \label{fig2}
\end{figure} 
In their seminal work \cite{WATANABE1994197}, Watanabe and Strogatz considered $N$ globally coupled phase oscillators having the following evolution equation:
\begin{equation}\label{eq1}
    \dot{\theta_j} = f + g\cos\theta_j + h\sin\theta_j, 
\end{equation}
where $j = 1,....,N$ and the coefficients $f,g,h$ must not depend on the index $j$. By considering the following transformation, WS theory showed that each trajectory is confined to a three dimensional subspace given by:
\begin{equation}
    tan\Big[\frac{1}{2} (\theta_j(t) - \Theta(t))\Big] = \sqrt{\frac{1 + \gamma(t)}{1 - \gamma(t)}} tan\Big[\frac{1}{2}(\psi_j - \Psi(t))\Big], \nonumber
\end{equation}
where $\psi_j$'s are constants and the entire dynamics is governed by the evolution of $\gamma(t), \Theta(t), \Psi(t)$. The systems obeying Eq.~\ref{eq1} can be reduced to the following closed form equations with three degrees of freedom and $N-3$ constants of motion:
\begin{equation}\label{eq2}
\begin{aligned}
    \dot{\gamma}       &= -\bigl( 1 - \gamma^2 \bigr) (g \sin\Theta - h \cos\Theta) \\
    \gamma \dot{\Psi}  &= -\sqrt{1 - \gamma^2} (g \cos\Theta + h \sin\Theta) \\
    \gamma \dot{\Theta} &= \gamma f - g \cos\Theta - h \sin\Theta.
\end{aligned}
\end{equation}
To avoid the singularity at $\gamma = 0$, Eq. ~\ref{eq2} can be written in the modified coordinates as follows:
\begin{equation}\label{eq3}
\begin{aligned}
  \dot{x} &= -yf + xyg + (1-x^2)h \\
  \dot{y} &= xf - xyh - (1-y^2)g \\
  \dot{\Phi} &= f - \frac{1 - \sqrt{1-\gamma^2}}{\gamma^2}(xg + yh),
\end{aligned}
\end{equation}
where $x = \gamma \cos\Theta$, $y = \gamma \sin\Theta$ and $\Phi = \Theta - \Psi$. This set of equation, has the complete information of dynamical evolution of the $N$ dimensional system (Eq.~\ref{eq1}) for all $N\geq3$. 
\begin{figure}[t!]
\centering
\includegraphics[scale=0.24]{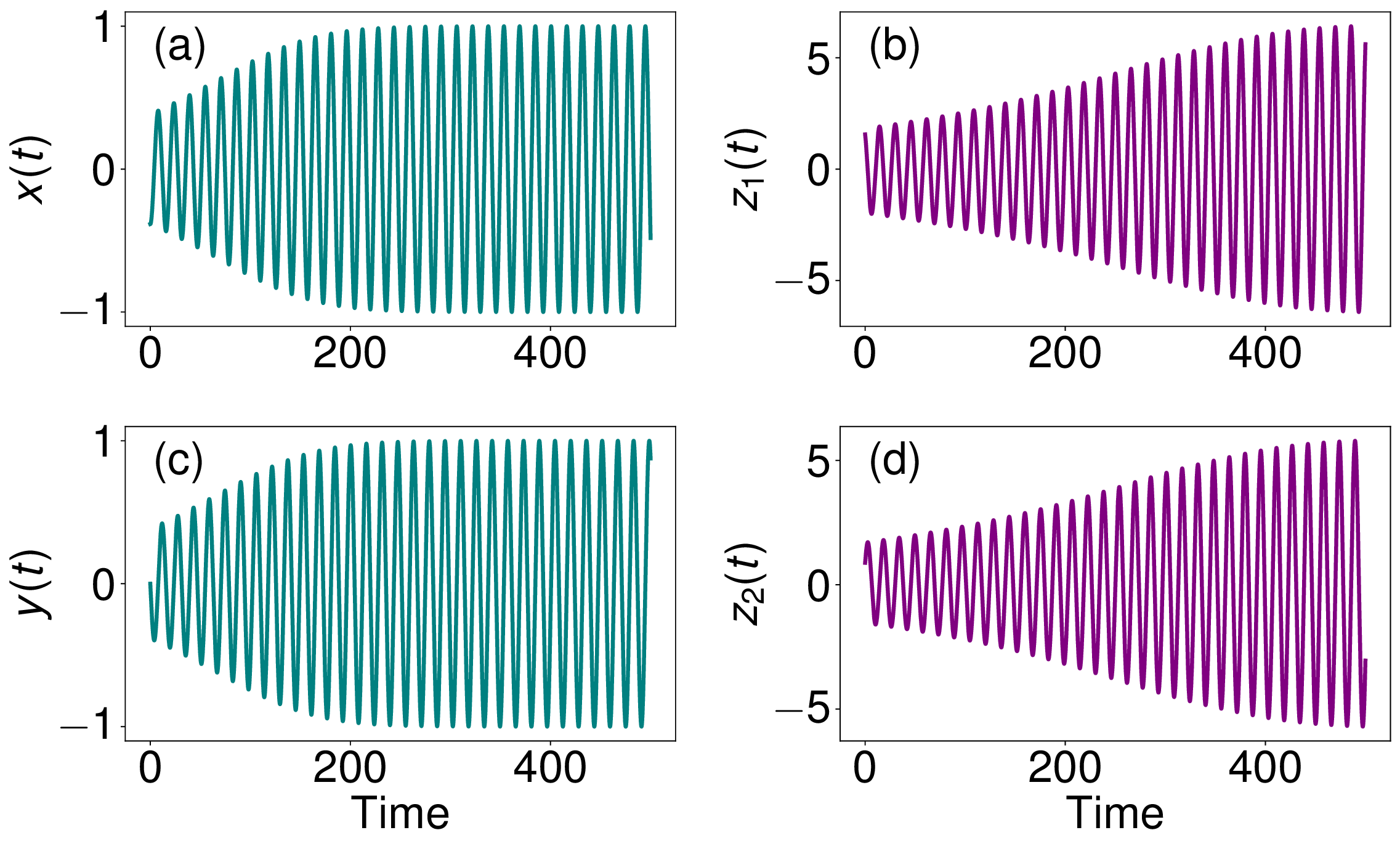}
\caption{Comparison of analytical WS variables and learned latent states. (a,\,c) Time evolution of analytical WS coordinates $x(t)$ and $y(t)$ for Eq.~\ref{eq5}. (b,\,d) Temporal trajectories of autoencoder latent variables $z_1(t)$ and $z_2(t)$.}
\label{fig3}
\end{figure}
The first prototype model we consider is a system of $N$ coupled Kuramoto oscillators with identical frequencies, whose governing microscopic dynamics can be rewritten directly in the mean-field form as:
\begin{equation}\label{eq4}
\begin{aligned}
    \dot{\theta}_i &= \omega + \frac{K}{N} \sum_{j=1}^{N} \sin(\theta_j - \theta_i) \\
    &= \omega + K r \sin\psi \cos\theta_i - K r \cos\psi \sin\theta_i,
\end{aligned}
\end{equation}
where the complex order parameter is defined as:
\begin{equation}
    r e^{\iota \psi} = \frac{1}{N}\sum_{j=1}^{N}e^{\iota\theta_j}.\nonumber
\end{equation}
Therefore, a direct comparison with Eq.~\ref{eq1} yields, $f = \omega$, $g = Kr\sin\psi$ and $h = -Kr\cos\psi$. The reduced system using Eq.~\ref{eq3} can be written as:
\begin{equation}
\label{eq5}
\begin{aligned}
    \dot{x}    &= -y \omega + K r \bigl( x y \sin\psi - (1 - x^2) \cos\psi \bigr) \\
    \dot{y}    &=  x \omega + K r \bigl( x y \cos\psi - (1 - y^2) \sin\psi \bigr) \\
    \dot{\Phi} &= \omega - K r \left( \frac{1 - \sqrt{1 - \gamma^2}}{\gamma^2} \right) (x \sin\psi - y \cos\psi).
\end{aligned}
\end{equation}
Given the initial conditions ($x(0), y(0), \Phi(0)$), the dynamics of the $N$-dimensional coupled system (Eq.~\ref{eq4}) are fully captured by Eq.~\ref{eq5}.
\begin{figure}[t!]
    \centering
    \includegraphics[width = \columnwidth]{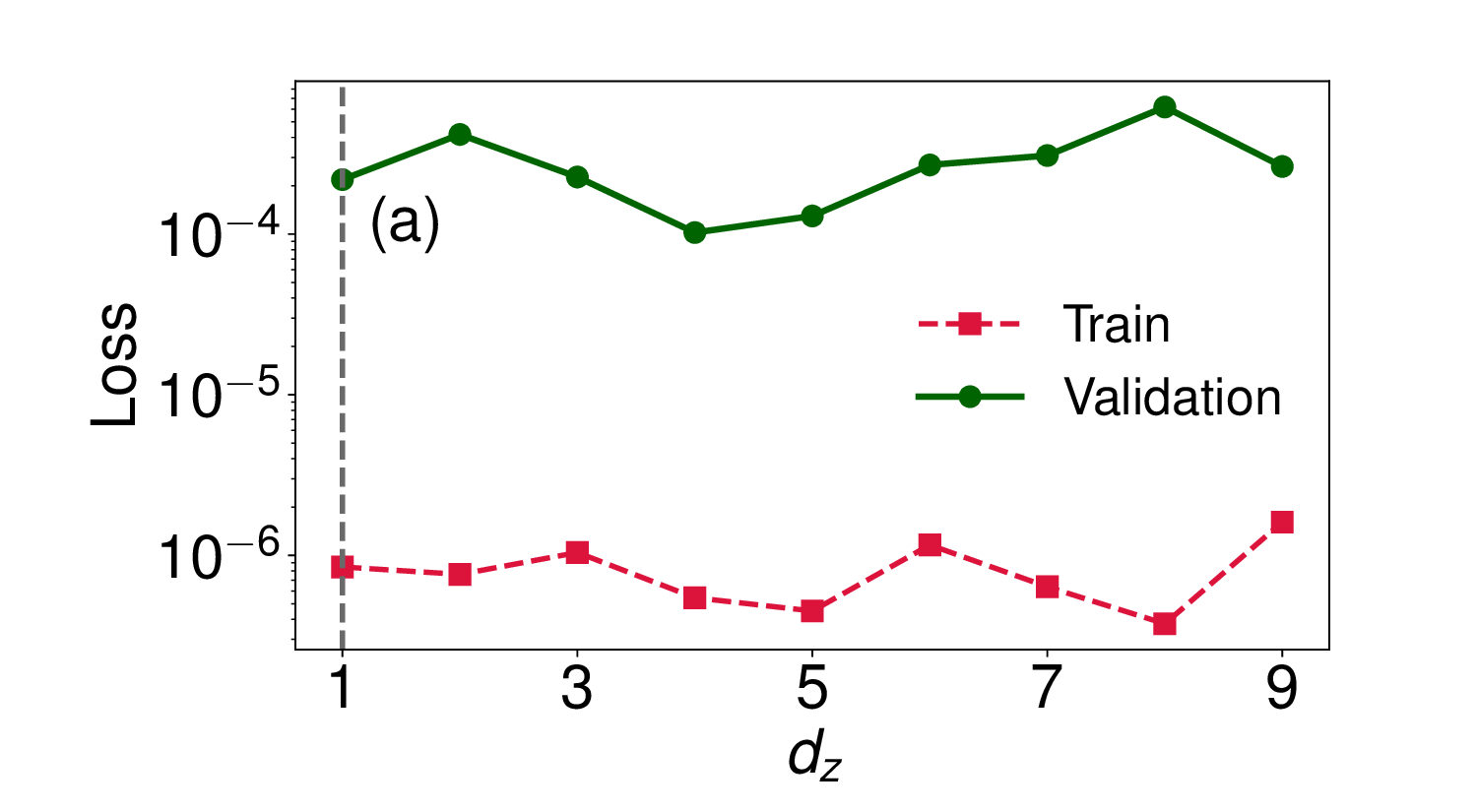} \\
    \includegraphics[width = \columnwidth]{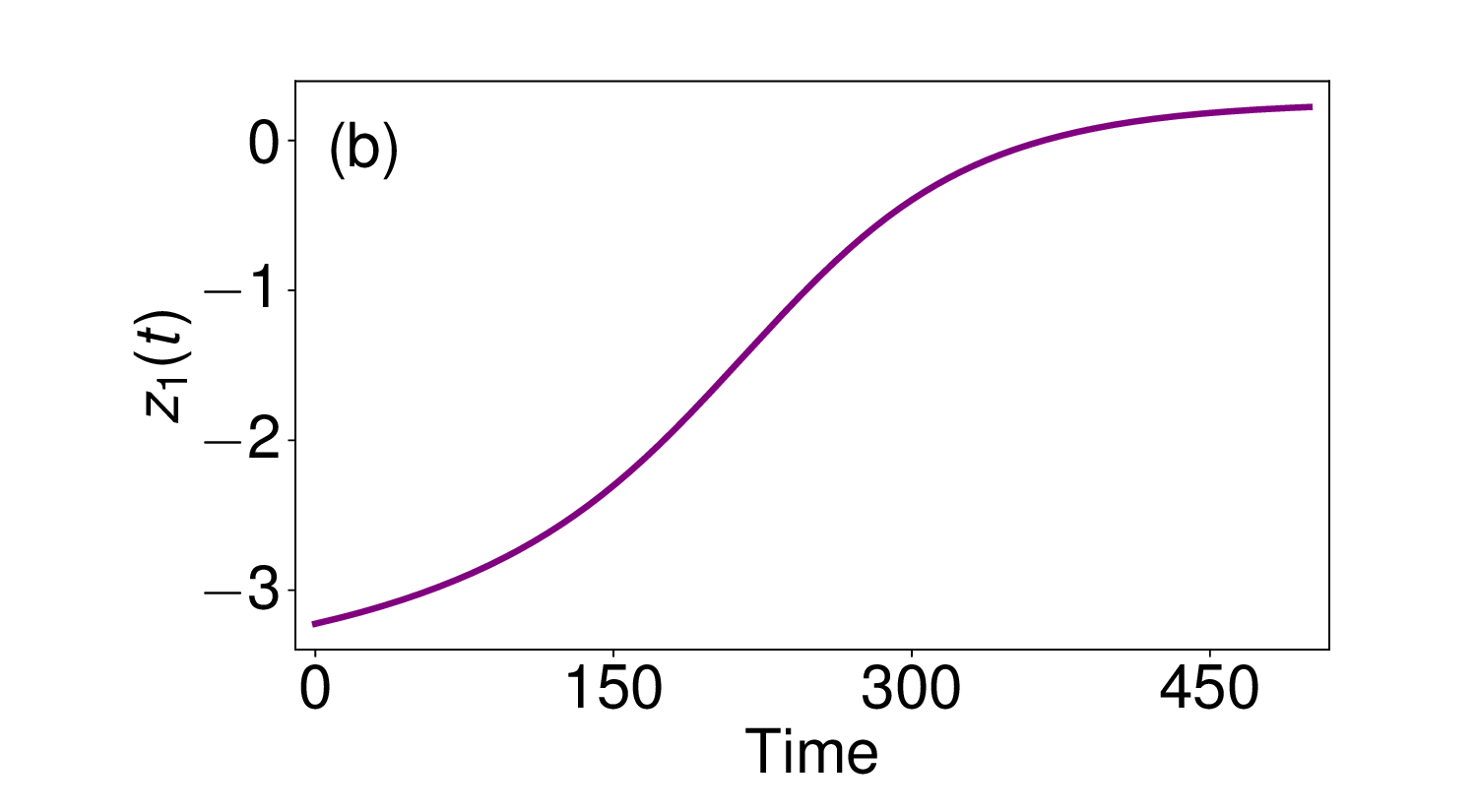}
    \caption{Identification of intrinsic dimensionality for $N = 20$ identical Kuramoto oscillators ($\omega = 0$) trained at coupling strength $K = 0.015$. (a) Training and MSE as a function of latent dimension $d_z$, exhibiting minimal error at $d_z = 1$ in agreement with the WS theory. (b) Temporal evolution of the learned latent variable $z_1(t)$, displaying non-oscillatory, monotonic growth that directly mirrors the analytical WS parameter $\gamma$. Training is performed with $B = 256$, $N_{\text{epochs}} = 200$.} 
    \label{fig4}
\end{figure}

As a second model, we examine an array of $N$ resistively loaded Josephson junctions, which similarly reduces to the governing dynamics in Eq.~\ref{eq1} in the mean field form:
\begin{equation}\label{eq6}
\begin{aligned}
    \dot{\theta}_i &= 1 + a\sin\theta_i + \frac{\epsilon}{N}\sum_{j = 1}^{N} \sin\theta_i \\
    &= 1 + a\sin\theta_i + \epsilon r \sin\psi.
\end{aligned}
\end{equation}
A direct comparison with Eq.~\ref{eq1} identifies the coefficients $f = 1+ \epsilon r\sin\psi$, $g = 0$ and $h = a$, leading to the following set of reduced equations:
\begin{equation}\label{eq7}
\begin{aligned}
    \dot{x}    &= -y \bigl( 1 + \epsilon r \sin\psi \bigr) + a (1 - x^2) \\
    \dot{y}    &=  x \bigl( 1 + \epsilon r \sin\psi \bigr) - a x y \\
    \dot{\Phi} &= 1 + \epsilon r \sin\psi - a y \left( \frac{1 - \sqrt{1 - \gamma^2}}{\gamma^2} \right),
\end{aligned}
\end{equation}
where the state evolution of the system governed by Eq.~\ref{eq6} is completely determined by the variables $x(t), y(t), \Phi(t)$ in Eq.~\ref{eq7}.
\section{Results}
\begin{figure}[t!]
    \centering
    \includegraphics[width = \columnwidth]{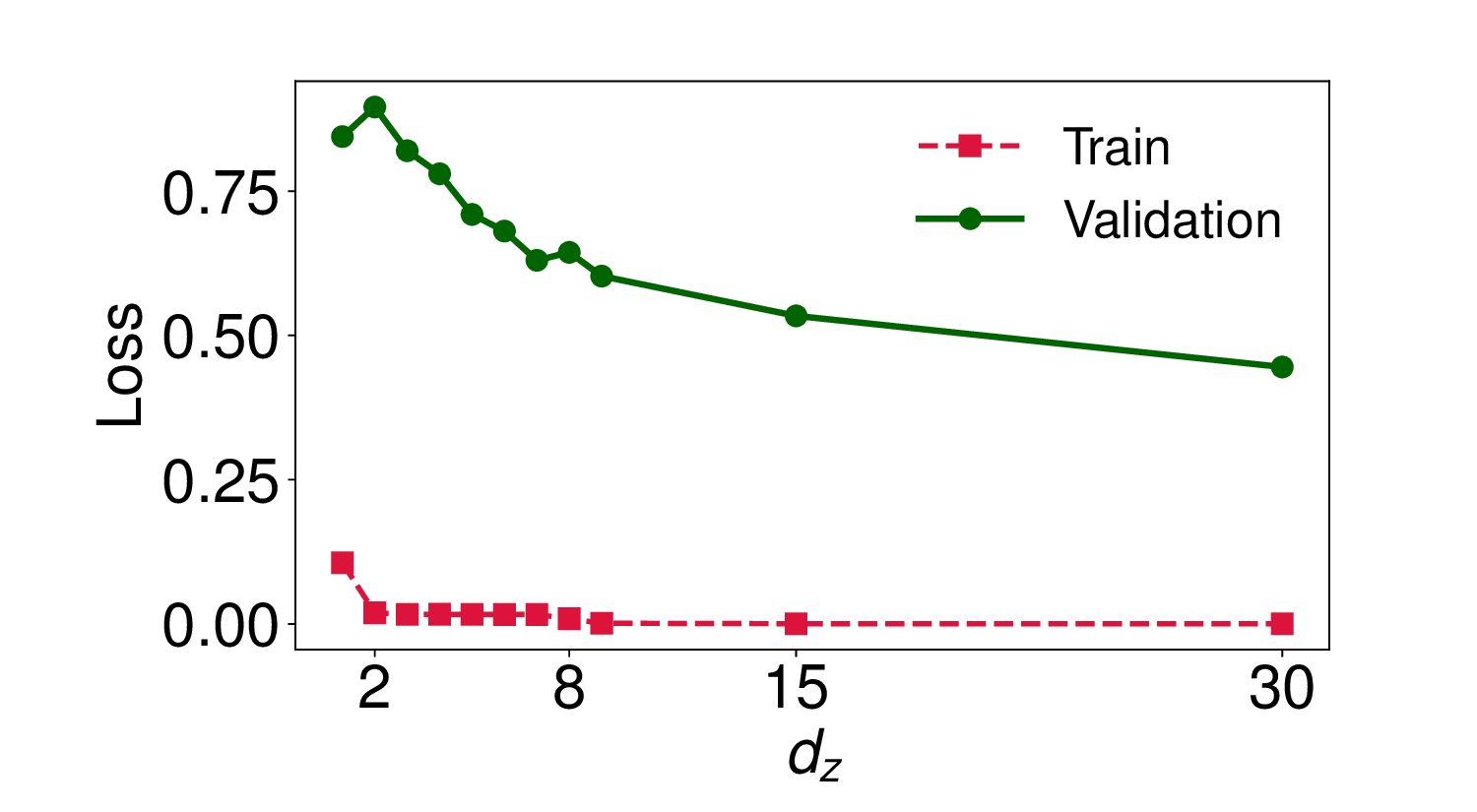} 
    \caption{Identification of intrinsic dimensionality under frequency heterogeneity. Globally coupled $N = 64$  Kuramoto oscillators with Lorentzian frequency distribution (mean $\omega_0 = 0$, FWHM ($\Gamma = 1$) trained at coupling strength $K = 1$. Unlike the identical case, the validation MSE exhibits a gradual decrease with increasing latent dimension $d_z$ rather than a sharp elbow transition. The pronounced gap between training and validation MSE reflects overfitting. Training is performed with $B = 64$, $N_{\text{epochs}} = 200$.}  
    \label{fig5}
\end{figure}
The autoencoder receives a $2N$-dimensional feature vector at each time step $t$, constructed by concatenating the trigonometric representations $[\cos\theta_j(t), \sin\theta_j(t)]_{j=1}^N$ of the phase trajectories. Training on this input sequence is carried out over $N_{\text{epochs}}$ epochs using the Adam optimizer with a mini-batch size of $B$. To determine the minimal latent dimension required to capture the dynamics, we systematically sweep the bottleneck size $d_z$. For each value of $d_z$, the network is independently reinitialized and trained, enabling us to quantify test-set reconstruction error as a function of latent dimensionality.
Under a change of variables $\gamma = \frac{2\rho}{1+\rho^2}$, $\Psi = \tilde{\Psi} + \pi$ and $\Theta = \tilde{\Theta} + \pi$, the WS transformation takes the form:
\begin{equation}\label{eq8}
    tan\Big[\frac{1}{2} (\theta_j(t) - \tilde{\Theta}(t))\Big] = \frac{1 - \rho(t)}{1 + \rho(t)} tan\Big[\frac{1}{2}(\psi_j - \tilde{\Psi}(t))\Big]
\end{equation}
The constants of motion $\{\psi_j\}_{j=1}^N$ are sampled uniformly on the circle, determining the initial oscillator phases $\theta_j(0)$ via Eq.~\ref{eq8}:
\begin{equation}\label{eq9}
    \theta_j(0) - \tilde{\Theta}(0) = 2 \tan^{-1} \left[ \frac{1-\rho(0)}{1 + \rho(0)} \tan \left( \frac{\psi_j - \tilde{\Psi}(0)}{2} \right) \right].
\end{equation}
To integrate the reduced WS dynamics (Eqs.~\ref{eq5}, \ref{eq7}) for direct comparison with the learned latent coordinates, we adopt the initial conditions $\tilde{\Theta}(0) = 0$, $\tilde{\Psi}(0) = 2\pi$, and $\gamma(0) = \frac{2\rho(0)}{1 + \rho^2(0)}$. Consequently, the initial transformed states evaluate to $x(0) = -\gamma(0)$, $y(0) = 0$, and $\Phi(0) = -2\pi$.

\subsection{Coupled Kuramoto oscillators}

\begin{figure}[t!]
    \centering
    \includegraphics[width = \columnwidth]{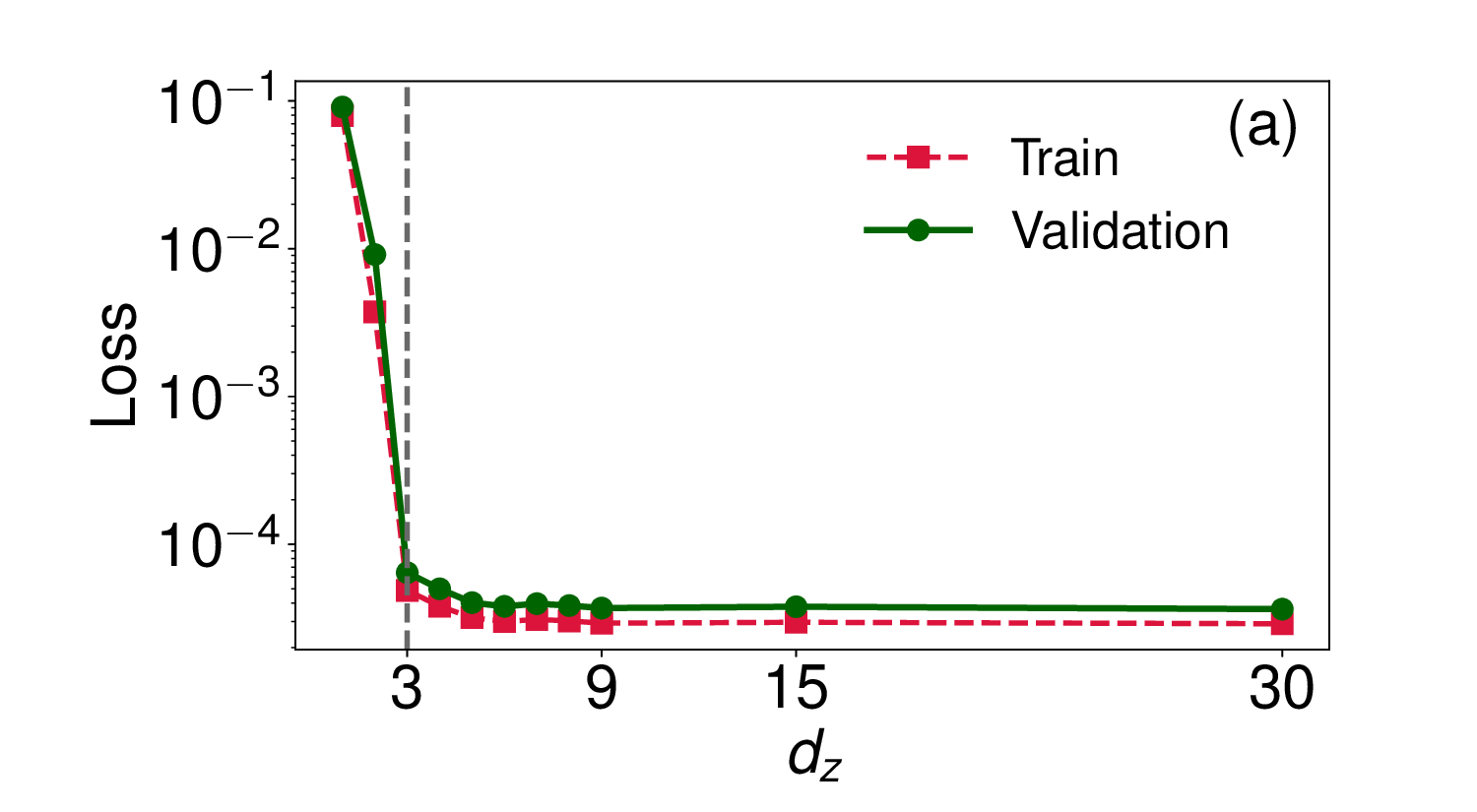} \\
    \includegraphics[width = \columnwidth]{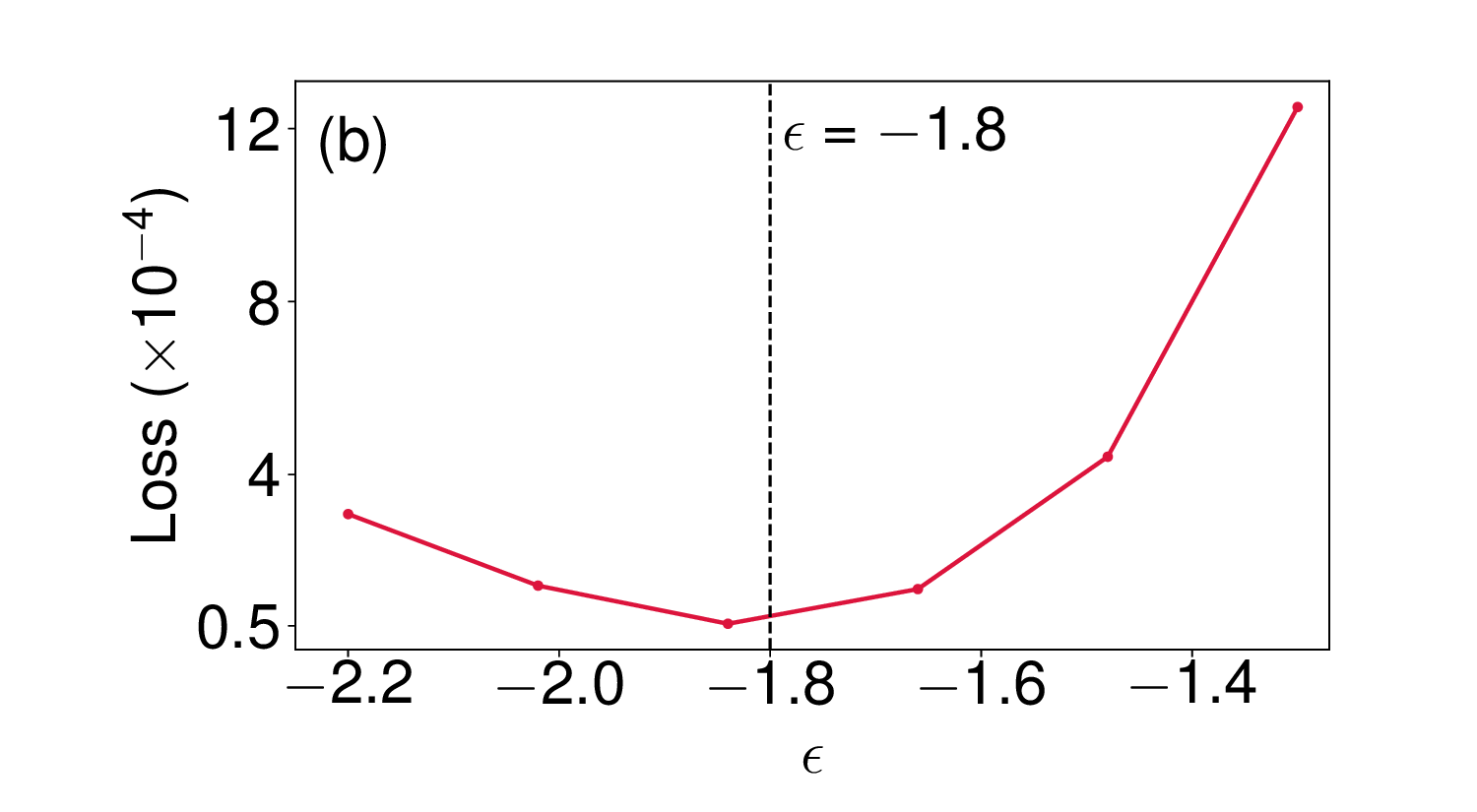}
    \caption{Identification of intrinsic dimensionality for resistive array of $N = 64$ Josephson Junctions ($a = 1$) trained at coupling strength $\epsilon = -1.8$. (a) Training and validation MSE as a function of latent dimension $d_z$, exhibiting a sharp elbow at $d_z = 3$ in exact agreement with the WS theoretical prediction. (b) MSE evaluated on unseen trajectories across varying coupling strengths $K$. The vertical dashed line marks the training parameter ($\epsilon = -1.8$), demonstrating generalization across different dynamical regimes. Training is performed with $B = 256$, $N_{\text{epochs}} = 350$}.
    \label{fig6}
\end{figure}
Phase trajectories $\theta_j(t)$ are generated by integrating Eq.~\ref{eq4} for $N = 64$ identical Kuramoto oscillators over $50{,}000$ time steps, using natural frequency $\omega = 0.4$, coupling strength $K = 0.015$, and initial amplitude $\rho(0) = 0.2$. To evaluate model performance, the initial and final $40\%$ segments of the trajectory are combined into the training set, while the central $20\%$ is held out as an unseen test set. In the thermodynamic limit ($N \to \infty$), the WS reduction converges to the Ott-Antonsen (OA) ansatz \cite{10.1063/1.2930766} when the constants of motion $\{\psi_j\}$ are uniformly distributed over the circle. In this continuum limit, the macroscopic WS variables map directly to the global complex order parameter \cite{PIKOVSKY2011872} via $\rho e^{\iota \tilde{\Theta}} = r e^{\iota \psi}$, with finite-$N$ deviations being exponentially small \cite{PIKOVSKY2011872}. Consequently, due to the resulting equivalence $\tilde{\Theta} = \psi$, the term $g\cos\Theta + h\sin\Theta = Kr(\sin\psi\cos\Theta - \cos\psi\sin\Theta)$ reduces to zero. Therefore, it can be inferred from Eq.~\ref{eq2} that $\dot{\Phi} = \dot{\Theta} = \omega$ and $\dot{\Psi} = 0$. Since $x = \gamma \cos\Theta$ and $y = \gamma \sin\Theta$, the $\dot{\Phi}$ equation becomes redundant, leaving $x,y$ as the two reduced variables. As illustrated in Fig.~\ref{fig2} a), the test reconstruction MSE drops sharply when the latent dimension is set to $d_z = 2$, beyond which additional dimensions yield negligible performance gains. This sharp elbow aligns directly with the WS theoretical framework since the reduced variables $(x, y)$ capture the complete dynamical state of the system. 

Moreover, since the analytical WS transformation (Eq.~\ref{eq8}) is independent of the coupling parameter, an autoencoder that successfully approximates this low-dimensional manifold representation should demonstrate generalization across different dynamical regimes. To test this, we evaluate the pre-trained architecture across an unseen range of coupling parameters ($K \in [0.001, 0.1]$) without retraining. As shown in Fig.~\ref{fig2} b), the reconstruction error remains bounded across the parameter sweep, with a minor increase in the error for the trajectories sampled far from the training parameter.
\begin{figure}[t!]
    \centering
    \vspace{0.5cm}
    \includegraphics[trim={1cm 0 0 0}, clip, scale=0.24]{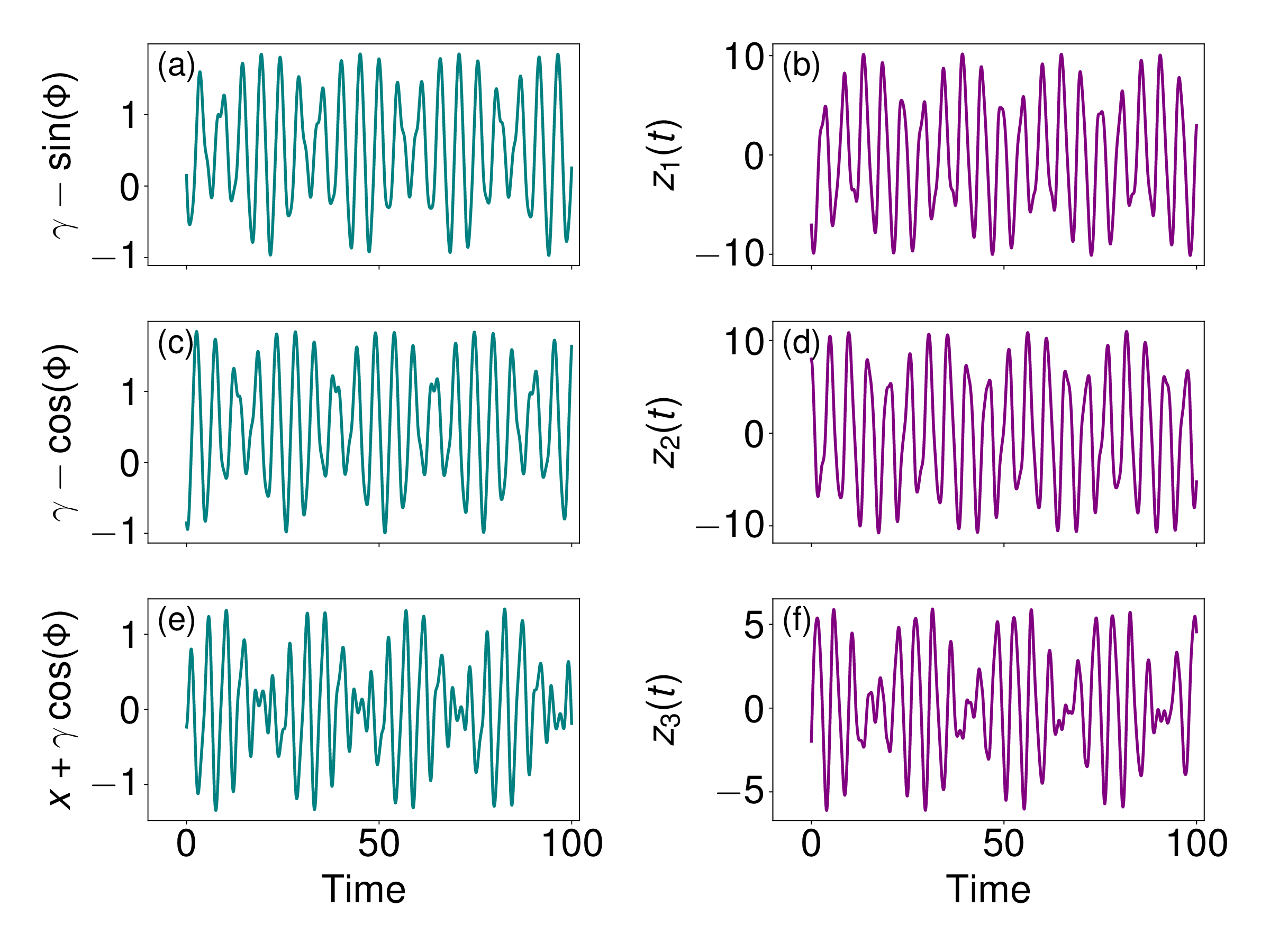}   
    \caption{Comparison of analytical WS variables and learned latent states. (a,\, c,\, e) Time evolution of heuristically constructed linear combinations of analytical WS variables ($\gamma - \sin\Phi$, $\gamma - \cos\Phi$, and $x + \gamma \cos\Phi$) of Josephson arrays (Eq.~\ref{eq7}), displayed for qualitative comparison. (b,\, d,\, f) Temporal evolution of the autoencoder latent variables $z_1(t)$, $z_2(t)$, and $z_3(t)$.}
\label{fig7}
\end{figure} 
To rigorously evaluate whether the latent space recovers these theoretical coordinates, we compare the learned latent variables $\mathbf{z}(t)$ with the WS variables (Fig.~\ref{fig3}). Beyond the apparent visual correspondence, we quantitatively validate this mapping by training a lightweight Multi-Layer Perceptron (MLP) regressor, comprising a single hidden layer with $N$ neurons and \textit{SiLU} activation, to predict the analytical WS variables $x(t), y(t)$ directly from the latent vector $\mathbf{z}(t)$. Employing the same evaluation strategy (training on the initial and final $40\%$ trajectory segments and holding out the central $20\%$), the regressor achieves a coefficient of determination of $R^2 = 0.9936$ alongside minimal validation loss. This high degree of predictability validates that the autoencoder's latent space is a nonlinear reparameterization of the WS variables and the information content of $(x,y)$ is preserved in $\mathbf{z}(t)$.

As a special limiting case with vanishing natural frequency ($\omega = 0$), the reduced system in Eq.~\ref{eq5} collapses further into a single effective dynamic equation. Since the global phase parameters become stationary ($\dot{\Phi} = \dot{\Theta} = 0$), the system is fully governed by the single remaining active variable, $\gamma$. Retraining the autoencoder on a array of $N = 20$ oscillators for $N_{\text{epoch}} = 200$, $B =256$ we observe that a single latent dimension ($d_z = 1$) is sufficient to achieve a dominant reduction in reconstruction error (Fig.~\ref{fig4} a). Furthermore, probing the temporal evolution of this single latent variable reveals a completely non-oscillatory trajectory (Fig.~\ref{fig4} b), mirroring the frozen global phase dynamics and reinforcing that the network faithfully recovers the underlying theoretical coordinate.

\subsection{Testing for irreducible dynamics}
Frequency heterogeneity breaks the continuous symmetry essential for exact low-dimensional WS reduction. Although the OA ansatz demonstrates that the asymptotic dynamics collapse onto a two-dimensional invariant manifold, this reduction holds strictly in the thermodynamic limit ($N \to \infty$). Furthermore, we specifically consider coordinate reductions that maintain a one-to-one mapping, enabling unique reconstruction of individual phase trajectories. Consequently, for any finite system size $N$, discrete finite-size fluctuations prevent exact manifold collapse, requiring the full $N$-dimensional state space to completely specify the dynamics.
To establish a control baseline and confirm that the sharp bottleneck MSE collapse are explicitly driven by exact WS reducibility, we repeat the experiment on a network of $N = 64$ non-identical Kuramoto oscillators at coupling strength $K = 1$ over $10{,}000$ time steps. Natural frequencies $\omega_i$ are made heterogeneous across units by sampling deterministically from a zero-mean Lorentzian distribution, with the full width half maxima (FWHM) $\Gamma = 1$:
\begin{equation}
    \omega_i = \tan \left( \frac{\pi (i - 0.5)}{N} - \frac{\pi}{2} \right). \nonumber
\end{equation}
For oscillator ensemble in the fully synchronized state, the system trivially reduces to a single collective phase variable. To avoid this degenerate state and ensure non-trivial dynamics, we choose $K < 2$, operating within the asynchronous regime \cite{10.1063/1.2930766}. Retraining the autoencoder for $N_{\text{epoch}} = 200$, $B =256$, the reconstruction error profile across latent dimensions no longer exhibits a sharp elbow (Fig.~\ref{fig5}). Rather, the validation MSE exhibits a gradual decrease with increasing latent dimension $d_z$ instead of a sharp elbow transition reflecting information loss while compression. Furthermore, the pronounced gap between the training and validation MSE reflects overfitting, as the autoencoder fits sample-specific trajectory fluctuations rather than isolating a true low-dimensional invariant manifold.

\subsection{Coupled Josephson Junctions}
To demonstrate the broader applicability of this data-driven framework, we extend our analysis to a second canonical system reducible to Eq.~\ref{eq1}. Numerical trajectories are generated for a network of $N = 64$ coupled resistive Josephson Junctions (Eq.~\ref{eq6}) with the parameter $a = 1$ and coupling strength $\epsilon = -1.8$, integrated over $30{,}000$ time steps. Identical to the previous setup, the constants of motion $\psi_j \in [0, 2\pi)$ are initialized uniformly on the unit circle, determining the initial phases $\theta_j(0)$ via Eq.~\ref{eq9} with $\rho(0) = 0.2$. To benchmark  the model performance on unseen trajectory intervals, the training set is composed of the initial $35\%$ and final $40\%$ segments, while the intermediate $25\%$ window is reserved for testing. Optimization is carried out for $N_{\text{epoch}} = 350$ epochs using mini-batch gradient descent with a batch size of $B = 64$. As depicted in Fig.~\ref{fig6} a), evaluating the validation MSE across varying bottleneck dimensions reveals a pronounced drop at latent dimension $d_z = 3$, beyond which performance gains plateau. This three-dimensional elbow precisely matches the theoretical prediction of the WS framework (Eq.~\ref{eq7}). Unlike the Kuramoto model, where one variable becomes redundant upon taking the constants of motion to be uniformly distributed, the Josephson Junction array lacks this collapse, necessitating the full state triplet $(x, y, \Phi)$ to uniquely specify the global dynamics. We evaluate the generalizability of the pre-trained architecture across unseen coupling regimes by sweeping the parameter $\epsilon  \in [-2.5, -0.75]$ without retraining network weights. As shown in Fig.~\ref{fig6} b), the reconstruction error remains well-behaved across moderate couplings but exhibits an asymmetric growth as $\epsilon \to 0$. This localized performance degradation likely stems from the onset of chaotic dynamics \cite{WATANABE1994197} near $\epsilon \to 0$, a dynamical regime absent from the training trajectory.

To verify whether the learned latent representation explicitly maps to the WS dynamics of the Josephson junction array, we compare the extracted coordinates against the analytical WS state variables (Fig.~\ref{fig7}). Visually, the temporal trajectories of the latent vector $\mathbf{z}(t) = (z_1, z_2, z_3)$ closely mirror linear combinations of the analytical variables; specifically $\gamma - \sin\Phi$, $\gamma - \cos\Phi$, and $x + \gamma \cos\Phi$. Quantitatively, a single hidden-layer MLP regressor ($N$ units, SiLU activation) trained on a fixed temporal validation split predicts the theoretical state vector $(x, y, \cos\Phi, \sin\Phi)$ directly from $\mathbf{z}(t)$ with high fidelity ($R^2 = 0.9987$, negligible MSE). Collectively, these results confirm that the autoencoder autonomously learns a smooth, nonlinear coordinate transformation of the exact 3D WS manifold in this system.

To evaluate the robustness and generality of our framework, we consider two additional systems exhibiting multi-population dynamics. While the single-population Kuramoto and Josephson junction array models analyzed in the sections above consist of identical oscillators, the WS reduction extends to heterogeneous ensembles by partitioning the system into $M$ coupled subpopulations of identical units \cite{PIKOVSKY2011872}. In this formulation, the microscopic dynamics map to a $3M$-dimensional set of governing WS equations (three per subpopulation). In the appendix ~\ref{sec:AppA} ,~\ref{sec:AppB}, we apply our framework to two extensions of Kuramoto models with $M = 2$ subpopulations.

\section{Conclusion}
In this work, we demonstrate an unsupervised machine learning framework to  discover low-dimensional macroscopic dynamics in coupled phase oscillator systems directly from trajectory data, without prior knowledge of analytical coordinate transformations. By unifying latent bottleneck diagnostics with nonlinear regression probing, the framework reliably captures the reduced state spaces predicted by WS theory across diverse physical systems. The autoencoder identifies the two-dimensional invariant manifold in identical Kuramoto oscillators (collapsing to a single macro-variable for vanishing natural frequencies), recovers the three-dimensional intrinsic dimensionality of resistive Josephson junction arrays, and extends seamlessly to multi-population architectures, including chimera states in two-subpopulation networks and contrarian–conformist dynamics. Moreover, latent space probing via MLP regression demonstrates that the learned representations form a smooth, nonlinear mapping to the analytical WS variables, preserving the full information of the theoretical coordinates. Furthermore, breaking exact WS reducibility via a Lorentzian frequency distribution works as a control experiment to confirm diagnostic specificity. For this case, as the Kuramoto dynamics does not admit an exact low-dimensional reduction, the sharp bottleneck elbow characteristic of reducible systems transforms into a monotonic loss curve. 

This framework highlights how deep neural networks can serve as quantitative tools for validating dimensional reductions and isolating effective physical degrees of freedom. Since the bottleneck latent representations are learned only as an approximation of the exact coordinate transform underlying the reduced dynamics, the extracted latent space remains a ``black box" lacking explicit analytical form. A vital direction for future investigation is establishing interpretability frameworks that analyze the trained autoencoder, enabling the direct recovery of explicit, closed-form coordinate transformations from its internal structure. Beyond validating known theories, this data-driven paradigm opens promising avenues for uncovering hidden symmetries and exact macroscopic reductions in complex, high-dimensional physical systems where analytical solutions are not yet known.

\begin{acknowledgments}
SJ gratefully acknowledges SERB Power grant award SPF/2021/000136. DS thanks CSIR fellowship under award no. JRF-NET2024/14347.
\end{acknowledgments}

\appendix
\setcounter{equation}{0}
\setcounter{figure}{0}
\renewcommand{\theequation}{\thesection.\arabic{equation}}
\renewcommand{\thefigure}{\thesection.\arabic{figure}}

\section{Chimera dynamics}\label{sec:AppA}
Networks of identical, symmetrically coupled oscillators can spontaneously split into coexisting coherent and incoherent domains, giving rise to chimera states \cite{PhysRevLett.101.084103}. This behavior is captured by a pair of oscillator populations in which each oscillator is coupled equally to all the others in its group, and less strongly to those in the other group.
\begin{equation}\label{eqA1}
    \dot{\theta}_{i}^{\sigma} = \omega + \sum_{\sigma' = 1}^{2}\frac{K_{\sigma \sigma'}}{N_{\sigma'}} \sum_{j=1}^{N_{\sigma'}}\sin\left(\theta_j^{\sigma'} - \theta_i^{\sigma} + \beta\right),
\end{equation}
where $\sigma \in \{1, 2\}$ labels the subpopulation and $N_{\sigma}$ is the corresponding population size. The intra-population coupling $K_{11} = K_{22} = \mu$ is stronger than the inter-population coupling $K_{12} = K_{21} = \nu$. Defining the complex order parameter for each subpopulation separately:
\begin{equation}
    r_{\sigma} e^{\iota \psi_{\sigma}} = \frac{1}{N_{\sigma}}\sum_{j=1}^{N_{\sigma}}e^{\iota\theta_j^{\sigma}}, \nonumber
\end{equation}
the system can be rewritten in the mean field form as:
\begin{equation}
\begin{aligned}
    \dot{\theta}_i^{(1)} &= \omega + \mu r_1 \sin(\psi_1 - \theta_i^{(1)} + \beta) + \nu r_2 \sin(\psi_2 - \theta_i^{(1)} + \beta)\\
    \dot{\theta}_i^{(2)} &= \omega + \mu r_2 \sin(\psi_2 - \theta_i^{(2)} + \beta) + \nu r_1 \sin(\psi_1 - \theta_i^{(2)} + \beta), \nonumber
\end{aligned}
\end{equation}
implying that each subpopulation has the form of Eq.~\ref{eq1} with the coefficients $f_1 = f_2 = \omega$, $g_1 = \mu r_1 \sin(\psi_1 + \beta) + \nu r_2 \sin(\psi_2 + \beta)$, $h_1 = -\mu r_1 \cos(\psi_1 + \beta) - \nu r_2 
\cos(\psi_2 + \beta)$ and $g_2 = \mu r_2 \sin(\psi_2 + \beta) + \nu r_1 \sin(\psi_1 + \beta)$, $h_2 = -\mu r_2 \cos(\psi_2 + \beta) - \nu r_1 
\cos(\psi_1 + \beta)$. Consequently, the complete macroscopic dynamics are captured by the 6-dimensional state vector $(x_1, y_1, \Phi_1, x_2, y_2, \Phi_2)$ described by (Eq.~\ref{eq3}). To circumvent the numerical singularities, we write the system of equations in the transformed coordinates described in Eq.~\ref{eq8} and the complete dynamics is given as:
\begin{subequations}\label{eqA2}
\begin{align}
    \dot{\rho}_\sigma &= \frac{1}{2}\left(1 - \rho_\sigma^2\right) C_\sigma  \\
    \dot{\tilde{\Theta}}_\sigma &= \omega_\sigma + \frac{1 + \rho_\sigma^2}{2\rho_\sigma} S_\sigma  \\
    \dot{\tilde{\Psi}}_\sigma &= \frac{1 - \rho_\sigma^2}{2\rho_\sigma} S_\sigma
\end{align}
\end{subequations}
where the driving fields $C_\sigma$ and $S_\sigma$ are given by
\begin{subequations}
\begin{align}
    C_\sigma &= \mu r_\sigma \cos(\psi_\sigma - \tilde{\Theta}_\sigma + \beta) + \nu r_{\sigma'} \cos(\psi_{\sigma'} - \tilde{\Theta}_\sigma + \beta) \nonumber \\
    S_\sigma &= \mu r_\sigma \sin(\psi_\sigma - \tilde{\Theta}_\sigma + \beta) + \nu r_{\sigma'} \sin(\psi_{\sigma'} - \tilde{\Theta}_\sigma + \beta). \nonumber
\end{align}
\end{subequations}
Numerical trajectories are generated for a network of $N = 128$ ($N_1 = N_2 = 64$) coupled oscillators (Eq.~\ref{eqA1}) with $\omega = 0$, $\beta = 0.1 - \pi/2$ and coupling strengths $\mu = 0.65, \nu = 0.35$, integrated over $2 \times 10^5$ time steps. The constants of motion $\psi_{j}^{\sigma} \in [0, 2\pi)$ are initialized uniformly on the unit circle for both the subpopulations, determining the initial phases $\theta_{j}^{\sigma}(0)$ via Eq.~\ref{eq9} with the initial conditions $\tilde{\Theta_1}(0)=\tilde{\Theta_2}(0) = 0$, $\tilde{\Psi_1}(0) = \pi$, $\tilde{\Psi_2}(0) = 2\pi$, $\rho_{1}(0) = 0.2, \rho_{2}(0) = 0.6$. After removing an initial transient of $1 \times 10^5$ time steps, the trajectory data is fed into the autoencoder. Model performance is evaluated by holding out three evenly spaced validation blocks of $10^4$ time steps each, using the remaining time series for training. Training is carried out for $N_{\text{epoch}} = 300$ with a batch size $B = 1024$.
\begin{figure}[t!]
    \centering
    \includegraphics[width = \columnwidth]{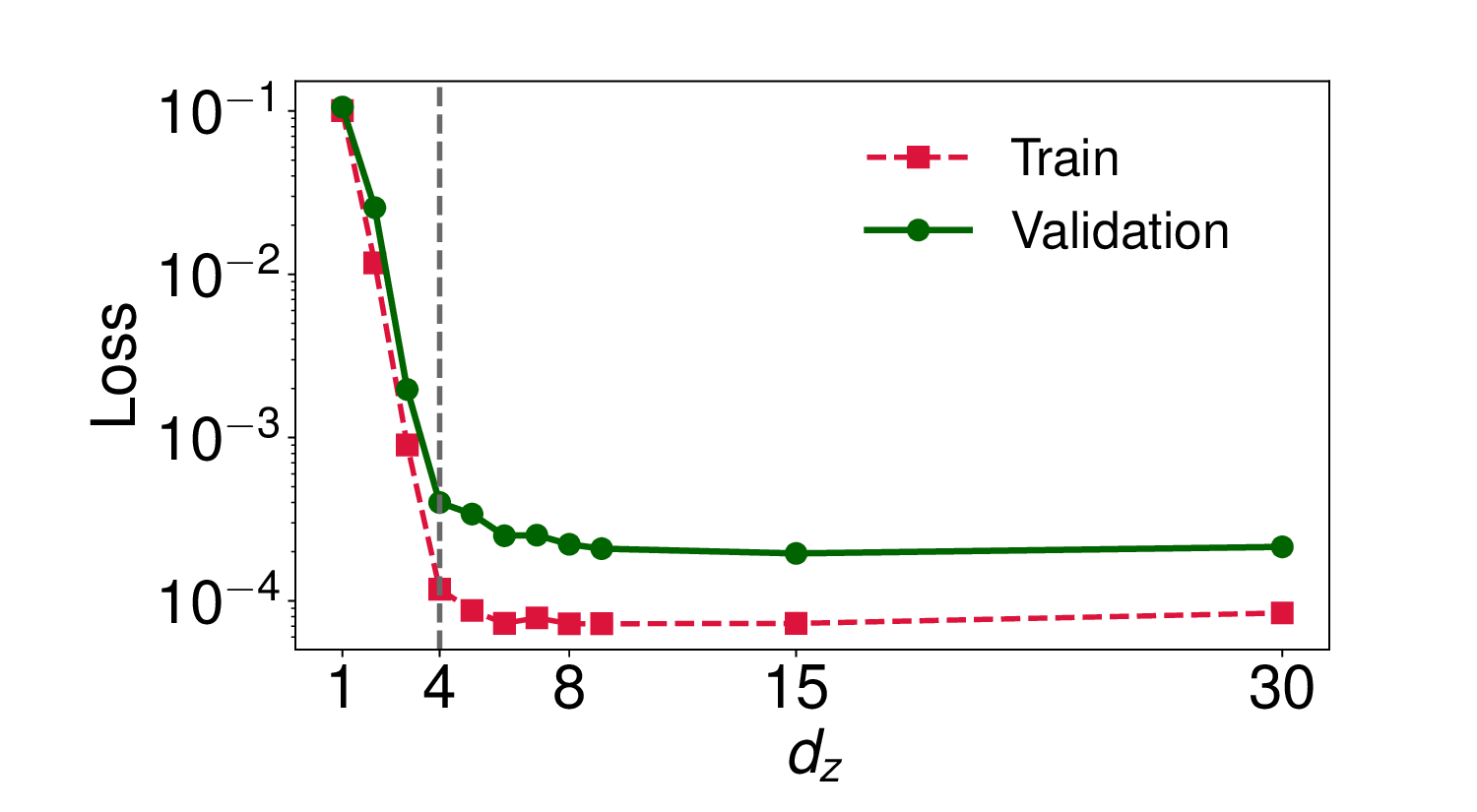} 
    \caption{Identification of intrinsic dimensionality for a model showing chimera. $N = 128$  Kuramoto oscillators with two subpopulations $N_1 = N_2 = 64$ having different intra-population ($\mu = 0.65$) and inter-population coupling ($\nu = 0.35$) trained at coupling strength $K = 1$. The validation MSE exhibits a pronounced drop at latent dimension $d_z = 4$ aligned with the WS theory. Training is performed with $B = 1024$, $N_{\text{epochs}} = 300$.}  
    \label{figA1}
\end{figure}
At the specified parameter values, the system enters a chimera state in which one subpopulation achieves complete synchronization ($\rho = 1$). Consequently, its dynamics collapse to $\dot{\rho} = \dot{\tilde{\Psi}} = 0$, leaving a single phase variable $\tilde{\Theta}$ to encode its macroscopic state [Eq.~\ref{eqA2}]. In contrast, the second subpopulation remains desynchronized with a time-varying order parameter \cite{PhysRevLett.101.084103}, necessitating the full triad of coordinates $(\rho, \tilde{\Theta}, \tilde{\Psi})$. Thus, the effective macroscopic manifold reduces from six to four dimensions to fully describe the system governed by Eq.~\ref{eqA1}. As depicted in Fig.~\ref{figA1}, the validation MSE shows a drop at the latent dimension $d_z = 4$, matching the theoretical prediction. A single hidden-layer MLP regressor trained in the same train-validation split to predict $\rho_{\sigma}, \cos\tilde{\Theta}_{\sigma}, \sin\tilde{\Theta}_{\sigma}, \cos \tilde{\Phi}_{\sigma}, \sin \tilde{\Phi}_{\sigma}$ directly from $\mathbf{z}(t)$ yields $R^2 = 0.9883$. These results further validate the non-linear mapping of the learned latent variables to the analytical WS variables.

\setcounter{equation}{0}
\setcounter{figure}{0}
\renewcommand{\theequation}{\thesection.\arabic{equation}}
\renewcommand{\thefigure}{\thesection.\arabic{figure}}

\section{Conformist and Contrarians model}\label{sec:AppB}
\begin{figure}[t!]
    \centering
    \includegraphics[width = \columnwidth]{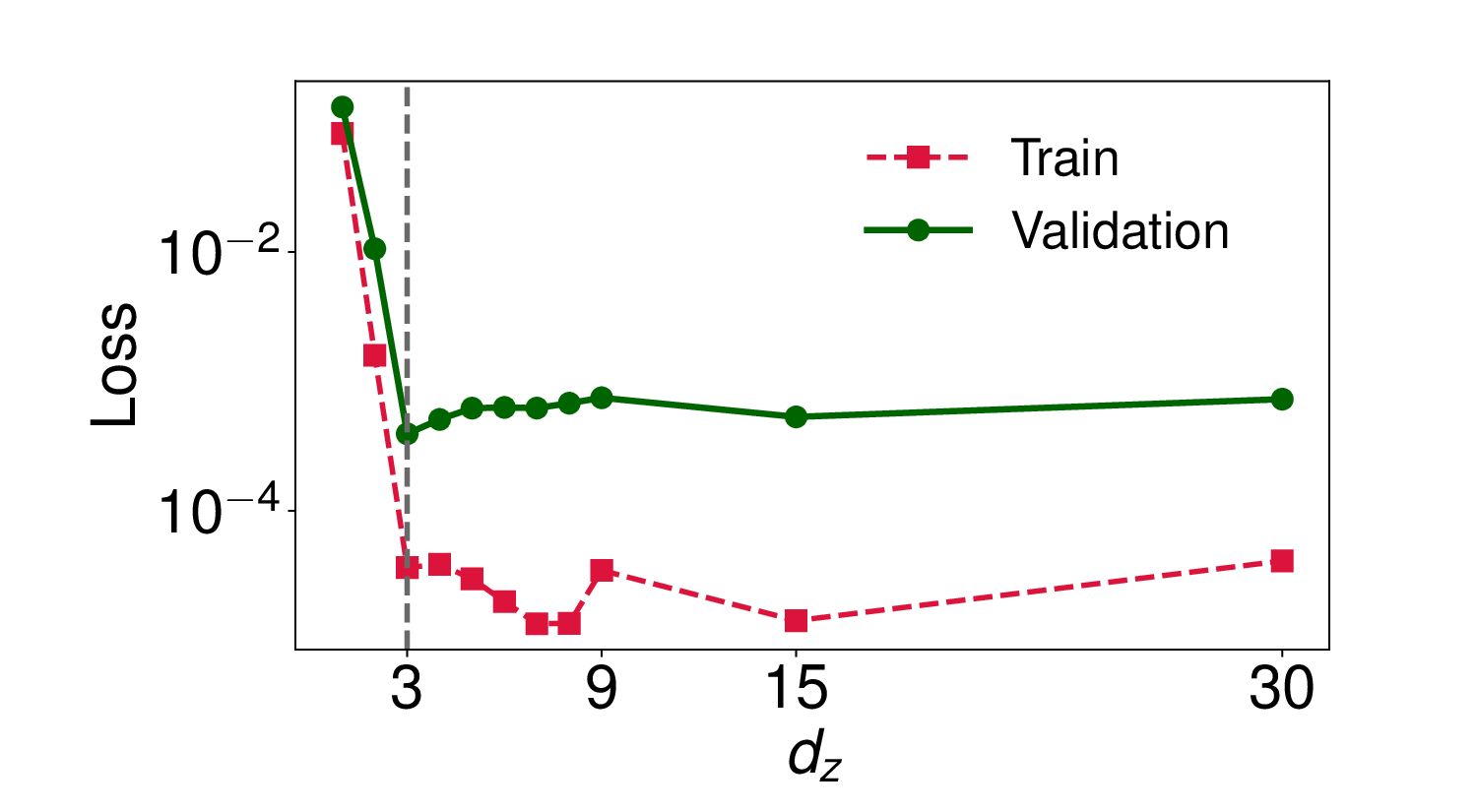} 
    \caption{Identification of intrinsic dimensionality for a contrarian-conformist model. $N = 64$  Kuramoto oscillators with the fraction of conformists $p = 0.52$, $\omega = 0$ and coupling values $K_1 = 1, K_2 = -0.5$. The validation MSE exhibits a pronounced drop at latent dimension $d_z = 3$ aligned with the WS theory. Training is performed with $B = 512$, $N_{\text{epochs}} = 300$.}    
    \label{figA2}
\end{figure}
The second model with two subpopulations is a generalization of Kuramoto model having both positive and negative couplings \cite{PhysRevE.84.046202}.
\begin{equation}\label{eqA4}
    \dot{\theta}_{i}^{\sigma} = \omega + \frac{K_{\sigma }}{N} \sum_{j=1}^{N}\sin\left(\theta_j - \theta_i^{\sigma}\right),
\end{equation}
where $\sigma \in \{1, 2\}$ labels the subpopulation and $N$ is the total population size. The oscillators in the first subpopulation are assumed to have positive coupling ($K_1 > 0$) to all the other oscillators in the system and are known as conformists, whereas, those in the second subpopulation have negative coupling ($K_2 < 0$) called contrarians. The fraction of conformists and contrarians is denoted by $p,q$ respectively. Defining the complex order parameter as:
\begin{equation}
    r e^{\iota \psi} = \frac{1}{N}\sum_{j=1}^{N}e^{\iota\theta_j}, \nonumber
\end{equation}
the system can be rewritten in the mean field form as:
\begin{equation}
\begin{aligned}
    \dot{\theta}_i^{(1)} &= \omega + K_1r\sin(\psi - \theta_i^{(1)}) \\
    \dot{\theta}_i^{(2)} &= \omega + K_2r\sin(\psi - \theta_i^{(2)}), \nonumber
\end{aligned}
\end{equation}
implying that each subpopulation has the form of Eq.~\ref{eq1} with the coefficients $f_1 = f_2 = \omega$, $g_1 = K_1r\sin\psi$, $g_2 = K_2 r\sin\psi$, $h_1 = -K_1r\cos\psi$ and $h_2 = -K_2r\cos\psi$. Consequently, the complete macroscopic dynamics are captured by the 6-dimensional state vector $(x_1, y_1, \Phi_1, x_2, y_2, \Phi_2)$ described by (Eq.~\ref{eq3}). Phase trajectories are simulated for a network of $N = 64$ ($p = 0.52$) coupled oscillators [Eq.~\ref{eqA4}] with zero natural frequency ($\omega = 0$) and coupling parameters $K_1 = 1$ and $K_2 = -0.5$, integrated over a total duration of $2 \times 10^5$ time steps. The constants of motion $\psi_{j}^{\sigma} \in [-\pi, \pi)$ are initialized uniformly on the unit circle across both subpopulations, specifying the initial phases $\theta_{j}^{\sigma}(0)$ via Eq.~\eqref{eq9} with the macroscopic initial state $\tilde{\Theta}_1(0) = \tilde{\Theta}_2(0) = 0$, $\tilde{\Psi}_1(0) = \pi$, $\tilde{\Psi}_2(0) = 2\pi$, and $\rho_{1}(0) = \rho_{2}(0) = 0.2$. After discarding an initial transient phase of $1.2 \times 10^5$ time steps, the remaining trajectory data are supplied to the autoencoder. Model performance is evaluated by holding out five uniformly spaced validation blocks of $5000$ time steps each, using the rest of the time series for training. The network is trained over $N_{\text{epoch}} = 300$ epochs with a batch size of $B = 512$.
Under this choice of parameters, the system settles into a traveling wave state where the conformist and contrarian subpopulations exhibit complete and partial phase synchronization, respectively. In accordance with Eq.~\eqref{eq2}, complete coherence in the conformist subpopulation sets $\gamma = 1$ ($\dot{\gamma} = \dot{\Psi} = 0$), leaving $\Theta$ as the sole dynamic coordinate required to describe its dynamics. Conversely, for the contrarian subpopulation, $\gamma$ attains a constant stationary value, so that two variables $(\Theta, \Psi)$ remain necessary to capture its dynamics. Therefore, the effective macroscopic manifold collapses from six to three dimensions for the system governed by Eq.~\eqref{eqA4}. As shown in Fig.~\ref{figA2}, the validation MSE drops sharply at latent dimension $d_z = 3$, aligning precisely with the theoretical prediction. A single hidden-layer MLP regressor trained on the same train-validation split to predict $x_{\sigma}, y_{\sigma}, \cos \Phi_{\sigma}$, and $\sin \Phi_{\sigma}$ directly from $\mathbf{z}(t)$ achieves $R^2 = 0.9972$. These results validate that the learned latent features capture the complete macroscopic WS dynamics.
\bibliography{references}

\end{document}